\documentclass[twocolumn,showpacs,prl,amsmath,amssymb]{revtex4}
\usepackage{graphicx}
\usepackage{dcolumn}
\usepackage{bm}
\usepackage{textcomp}

\begin{document}
\title{Frequency shift of optical phonons in doped graphene layers}

\author{Srijan Kumar Saha$^1$, U. V. Waghmare$^2$,
H. R. Krishnamurthy$^{1,2}$ and A. K. Sood$^1$}
\affiliation{$^1$Department of Physics, Indian Institute of Science,
 Bangalore 560012, India\\
$^2$Theoretical Sciences Unit,
Jawaharlal Nehru Centre for Advanced Scientific Research,
Bangalore 560064, India}
\date{\today}

\begin{abstract}
We use first-principles density-functional calculations to determine the frequency shift 
of the A$'_1$-${\bf K}$ phonon (Raman D band) in monolayer graphene, as a 
function of the charge doping. A detailed DFT study on the electron-phonon 
coupling and the phonon line width of E$_{2g}$-${\bm \Gamma}$
 and A$'_1$-${\bf K}$ phonons are also performed for
graphene multi-layers. Furthermore, we explain the experimentally observed 
'1/(Number of Layers)' behaviour of the 
Raman G band position after including the dynamic response treated 
within time dependent perturbation theory.    
\end{abstract}
                 
\maketitle

Graphene, a two-dimensional honeycomb lattice of $sp^2$-bonded carbon atoms~\cite{r1}, is known to exhibit 
many remarkable properties that are of fundamental interest and technological relevance. Being
 the fundamental building block for carbon allotropes of other
dimensionality, it can be stacked into 3d graphite or rolled into 1d nanotube. 
Graphene has created a great interest in the 
scientific community due to its excellent mechanical and electronic 
characteristics, scalability to nanometer sizes~\cite{r2}
 and easy accessibility to
optical probes. Moreover, this atomically-thin sheet is thermodynamically stable, 
continuous on a macroscopic scale and exhibits high crystal quality. A very
intriguing feature of graphene is related to the fact that its electron 
transport is essentially governed by the Dirac's (relativistic) equation and 
can be controlled externally~\cite{r3,r4,r5,r6}. Using electric field in an FET geometry
 it is possible
 to dope graphene layers by changing the carrier concentration in the 
samples.
In particular, its phonon spectrum can be modified significantly  by tuning the applied
gate voltage~\cite{r7,r8}.

Phonon dispersions of graphene exhibit two Kohn anomalies in the highest optical
branches at ${\bm \Gamma}$ and ${\bf K}$ (E$_{2g}$ mode - 
Raman G band and A$'_1$ mode - Raman D band, respectively).
The electron phonon couplings   
for these  modes are particularly large and negligible for all the other modes
at ${\bm \Gamma}$ and ${\bf K}$~\cite{r9}. In this paper, 
we calculate the frequency 
shift of the long wavelength optical phonon (E$_{2g}$-${\bm \Gamma}$ mode)
in graphene as a function of the number of layers.
Phonon line width and the electron phonon coupling (EPC) for the 
above two optical phonons are also extensively studied in 
graphene multilayer. 
 Furthermore, for the first time, we
 compute the variation of phonon frequency of the 
 A$'_1$ mode at ${\bf K}$ in a graphene monolayer, as a function of doping
concentration. Calculations are done first using a fully ab-initio approach
within the standard Born-Oppenheimer approximation and then 
time-dependent perturbation theory is used to explore the effect of dynamic
response.

Our ab-initio calculations are performed using density functional 
theory (DFT)~\cite{r10}
in the generalized gradient approximation. We have used the PWSCF~\cite{r11}
implementation of DFT, with Perdew-Burke-Ernzerhof (PBE)~\cite{r12}
 for the exchange correlation functional and ultrasoft pseudo potential
~\cite{r13} to represent the interaction between ionic cores and valence
 electrons. Kohn-Sham wave functions are represented with a plane wave basis 
 truncated at 
 an energy cut off of 40 Ry. The two-dimensional graphene crystal is 
simulated using a supercell geometry with a vacuum of 7.5~\AA~ in the z-direction
  to ensure 
negligible interaction between its periodic images. The Brillouin zone
 integration is done on a uniform 
 $36\times 36\times 1$ Monkhorst-Pack~\cite{r14} 
 grid. An electron smearing of 0.02 Ry with Fermi-Dirac distribution is used
 to accelerate the slow convergence of self-consistent calculation in 
 graphene layers by smoothing out the discontinuities present in the 
 Fermi distribution at zero temperature.
 Structural relaxation is carried out in each case to minimize the forces 
acting on each of the atoms using Broyden-Flecher-Goldfarb-Shanno (BFGS) 
based method~\cite{r15}.
Phonon frequencies and electron phonon couplings are calculated using Density Functional Perturbation Theory (DFPT)~\cite{r16} at the level of linear response, which
allows the exact (within DFT) computations of phonon frequencies at any 
Brillouin zone point. The Fermi-energy shift  with doping is simulated by considering
an excess electronic charge which is compensated by a uniformly charged
back-ground.

In graphene, the Fermi surface is reduced to two equivalent         
 ${\bf K}$ and ${\bf K}$$'$ points in opposite corners of the 2D hexagonal
Brillouin zone where the valence and conducting bands touch each other. 
This  leads
to the Dirac cone spectrum 
$\epsilon_{\pi^*/\pi}({\rm \bf K}+{\bf k})=\pm\hbar v_F k$
for the $\pi^*$ and $\pi$ bands, where $v_F$ is the Fermi velocity of the massless
(2+1) dimensional Dirac fermions and ${\bf k}$ is a small vector.
Within this approximation, at zero temperature ($T$ = 0 K), the
layer charge concentration and the Fermi energy are related via  
\begin{eqnarray}
\sigma = {\rm sign}(\epsilon_F)~ 
\frac{\epsilon_F^2}{\pi(\hbar v_F)^2} 
 = {\rm sign}(\epsilon_F)~ 
{\epsilon_F^2}~ \nonumber \\
 \times~ 10.4~\times 10^{13}~ {{\rm cm}^{-2}}~{\rm (eV)}^{-2}
\label{eq1}
\end{eqnarray}
where $\hbar v_F=5.53$~eV$\cdot$\AA~from our DFT calculation,
in excellent agreement with the previous DFT result~\cite{r17} 
and $\epsilon_F=0$ at the $\pi$ bands crossing (known as Dirac point).
For very small doping concentration (order of $10^{13} cm^{-2}$), the linearized
 bands are a good approximation. We have used this model of strongly
 interacting two-dimensional  Dirac fermions 
 in few layer- graphene as an 
 approximation because the Dirac singularity is the topological property of the
 electronic spectra that should be stable towards the weak 3d inter-layer 
coupling even in few layer- graphene. That's why the Fermi velocity for the few 
layers of graphene~\cite{r18} 
 is taken to be the same as for a  single layer within
a good approximation for low doping concentration. To check
validity of the above approximation in
graphene multi-layers, we have performed the DFT calculations of the
electronic band structure for monolayer, bilayer, trilayer and tetralayer
graphene and our 
results ( Figure 1 ) confirm that the above assumption is justified. 

\begin{figure} [h]                                                                 \centerline{\includegraphics[width=85mm]{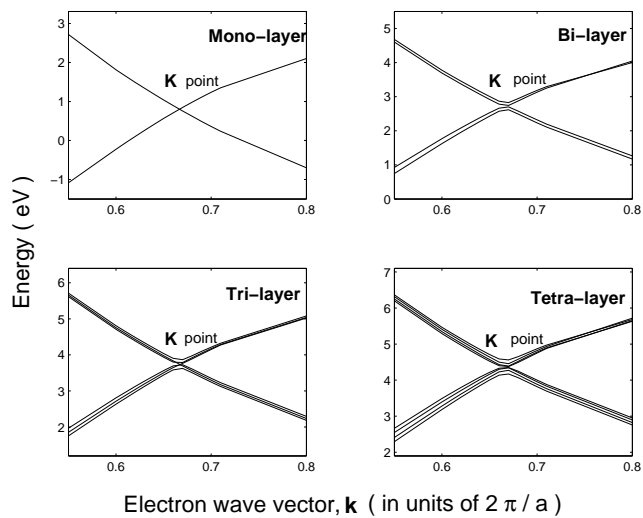}}                           \caption{                                                                          The electronic band structure of graphene layers
determined with ab-initio DFT calculations.}                                                                                  \label{fig1}                                                                       \end{figure} 

We first determine the total 
energy and electronic density by solving the self-consistent Kohn-Sham
equations for the equilibrium crystal geometry where the relative atomic 
positions in the unit cell yield zero forces and lattice parameters lead to
a zero stress-tensor.
Once the unperturbed ground state is determined, phonon frequencies are
 obtained by using DFPT which calculates
the linear response of the electrons to a static perturbation induced
by ionic displacements. This approach is based on the adiabatic (also referred
to as   
the Born-Oppenheimer or static) approximation.
At zero-doping, our computed E$_{2g}$-${\bm \Gamma}$
 and A$'_1$-${\bf K}$ 
 phonon frequencies in a monolayer graphene are $\omega_{\bm \Gamma}^{static} (0) = 1553.3$~cm$^{-1}$ 
and $\omega_{\bf K}^{static} (0) = 1301.7$~cm$^{-1}$,
respectively.
Infact, the phonon exactly at the ${\bf K}$ point is not the relevant
one for the Raman D or 2D bands in graphene; rather what is relevant
is the phonon at $({\bf K}$ $-\Delta$${\bf K})$ where $\Delta$ ${\bf K}$ is
determined by the double resonance process~\cite{r19, r20, r21}.
For the typical experimentally used laser excitation of 2.41 eV, 
the value of $\Delta {\bf K}$ is $0.0855$ (in units of $2 \pi/a$).
Our results of
the frequency shifts for the {\bf K} point phonon
($\Delta\omega_{\bf K}^{static}$) and the $({{\bf K}- \Delta {\bf K}})$ 
point phonon 
($\Delta\omega_{{\bf K}- \Delta {\bf K}}^{static}$) 
 with charge doping $\sigma$ 
are shown in Figure 2. In both panels, the open circles 
show the results obtained when the lattice spacing is kept fixed at 
its value corresponding to the undoped case and the filled stars 
show the results obtained when the lattice spacing is optimized
for each doping concentration. The optimization leads to a lattice 
expansion 
with electron doping and a lattice  contraction with hole doping, and gives 
 substantial change in the phonon frequency shift. 
Interestingly, Figure 2 shows that the frequency shift of
the A$'_1$-${\bf K}$ phonon on
the charge doping in a graphene monolayer is very different compared to the
${\bm \Gamma}$ point phonon~\cite{r17}.  

\begin{figure} [h]
\centerline{\includegraphics[width=85mm]{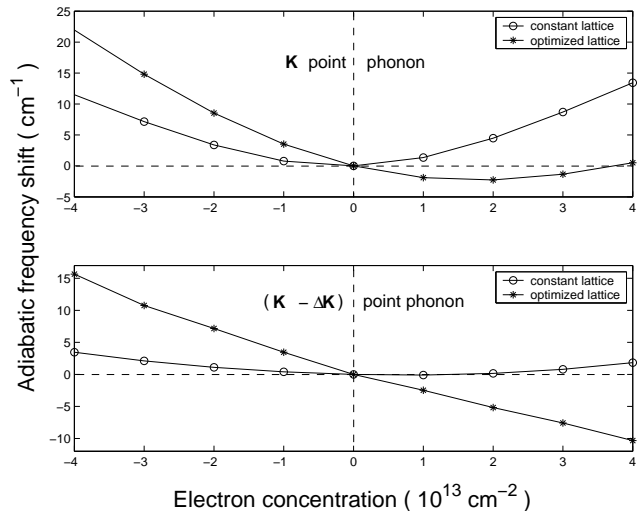}}
\caption{Adiabatic frequency shift of the ${\bf K}$-
 A$'_1$  and ${{\bf K}- \Delta {\bf K}}$ 
 phonon in graphene monolayer 
as a function of charge doping $\sigma$,
 with respect to the zero-doping frequency. 
}
\label{fig2}
\end{figure}

Within the adiabatic approximation, above results are generically expected 
on physical grounds. For the long-wavelength ${\bm \Gamma}$ phonon
 (small wave-vector
${\bf q}$), a high density of electron doping leads to an effective 
screening of the ion-ion interactions, reducing the elastic coupling 
in the lattice and hence leading to a softening of the phonons.
On the other hand, in the case of the
${\bf K}$-point phonon with a larger wave-vector 
which connects two points in the
Fermi surface, electrons are no longer able to effectively shield 
the ion-ion interactions and the vibrational properties shows a
giant and sharp Kohn-anomaly - a precipitous phonon softening at 
${\bf K}$ point. With doping, the change in the Fermi surface moves that
Kohn-anomaly away from ${\bf q}={\bf K}$ and stiffens ${\bf K}$-point
 phonon (Figure 2).

So far we have used the standard adiabatic approximation which is 
usually justified in ordinary metals where the electronic energy 
gap between the ground and the excited states is 
larger than the phonon energy so that the phonons respond to a time
averaged electron distribution. In graphene, however, this approximation
is inadequate because of its unique massless Dirac-like electron 
band dispersion and comparable electronic and phonon energy scale.
The complete  break-down of the adiabatic approximation 
in doped graphene has been
recently demonstrated, both theoretically and experimentally~\cite{r17,r7}.  
Hence, for single-, double- and few layer- doped graphene, we next 
consider the phonon as a 
dynamic perturbation
treated within time-dependent perturbation
theory. 

Using such a dynamic approach in the context of DFT, the 
self-energy of a phonon mode  $\nu$  at wave-vector ${\bf q}$ can be
written as~\cite{r22},
\begin{equation}
S_{\nu}({\bf q},\omega_{{\bf q}\nu})=\frac{2}{N_k}\sum_{{\bf k},i,j} |g_{({\bf k+q})i,{\bf k}j}^{\nu}|^2\,
\frac{f_{({\bf k}+{\bf q})i} - f_{{\bf k}j}}{\epsilon_{({\bf k}+{\bf q})i}-\epsilon_{{\bf k}j}-\hbar \omega_{{\bf q}\nu}-i\eta} 
\label{eq:PIdef}
\end{equation}
where $N_k$ is the number of ${\bf k}$-points, the sum is over ${\bf k \in}$ Brillouin zone, 
$f_{{\bf k}j}$ is the Fermi distribution function,  $\eta$ is a small
real number and $g_{({\bf k+q})i,{\bf k}j}^{\nu}$  is the electron-phonon matrix
element.
Within DFPT, the electron phonon matrix element can be obtained from the first
order derivative of the self-consistent Kohn-Sham potential, 
with respect to the atomic displacements as: 
$g_{({\bf k+q})i,{\bf k}j}^{\nu}= \langle ({\bf k+q})i|\delta V_{KS}/\delta u_{{\bf q}\nu} |{\bf k}j\rangle /\sqrt{2M \omega_{{\bf q}\nu}}$ where $u_{{\bf q}\nu}$ is the amplitude of the displacement of the phonon $\nu$
of wave vector ${\bf q}$,
$\omega_{{\bf q}\nu}$ its phonon frequency, M is the atomic mass, $V_{KS}$ the Kohn-Sham potential
and $|{\bf k}j\rangle$ is a Bloch eigenstate with wave-vector ${\bf k}$, 
band index j and energy $\epsilon_{{\bf k}j}$.

The dynamic correction to the energy shift 
arising from the electron-phonon interaction is 
given by the real part of the phonon self-energy as:
\begin{equation}
\frac{\hbar\Delta\omega_q}{2} = \frac{2}{N_k}\sum_{{\bf k},i,j} |g_{({\bf k+q})i,{\bf k}j}^{\nu}|^2\,
{\cal P}\left[\frac{f_{({\bf k}+{\bf q})i} - f_{{\bf k}j}}{\epsilon_{({\bf k}+{\bf q})i}-\epsilon_{{\bf k}j}-\omega_{{\bf q}\nu}} \right]
\label{RePI}
\end{equation} 
where ${\cal P}$ takes the principal part of its argument.

The phonon line width (FWHM) is twice the imaginary part of $S_{\nu}({\bf q},\omega_{{\bf q}\nu})$
and  can also be obtained from the Fermi golden rule~\cite{r23}:
\begin{eqnarray}
\gamma_{{\bf q}\nu}&=&\frac{4\pi}{N_k}
\sum_{{\bf k},i,j} |g_{({\bf k+q})i,{\bf k}j}^{\nu}|^2  \nonumber \\
&&\left(f_{{\bf k}j} - f_{({\bf k}+{\bf q})i}\right)
 \delta(\epsilon_{({\bf k}+{\bf q})i}-\epsilon_{{\bf k}j}-\omega_{{\bf q}\nu})
\label{ImPI} 
\end{eqnarray} 

Ideally, in the dynamic case, $\omega$ should be determined self-consistently.
However, considering the dynamic and doping effects as perturbation, at the
lowest order, we can use the adiabatic undoped phonon frequency $\omega_{0}^{static}$
in place of $\omega$.
For ${\bf q}$ near ${\bm \Gamma}$-point, we can also consider
the electron phonon coupling as
$|g_{({\bf K+k})j,({\bf K+k})i}|^2=\langle g^2_{\bm \Gamma}\rangle_{\rm F} [1\pm\cos(2\theta)]$,
where $\theta$ is the angle between the phonon-polarization ${\bf q}$ and ${\bf k}$,
the sign $\pm$ depends on the transition and $\langle g^2_{\bm
\Gamma}\rangle_{\rm F} = \sum_{i,j}^\pi |g_{{\rm \bf K}i,{\rm \bf
K}j}|^2/4$, the sum is for the two degenerate
$\pi$ bands at $\epsilon_{\rm F}$
(see Eqn.~6 and note 24 of Ref.~\cite{r9}).

Based on the above approximations with 
the linear electron dispersion near Dirac point,
we find that the change of Raman G-band phonon energy at zero
temperature can be described as,
\begin{equation}
\hbar\Delta\omega_{\bm \Gamma}^{dynamic}=
\alpha_{\bm \Gamma} |\epsilon_F| +
\frac{\alpha_{\bm \Gamma}\hbar\omega_{\bm \Gamma}^{static}(0)}{4}  
\ln \left|
\frac{2|\epsilon_F|-\hbar\omega_{\bm \Gamma}^{static}(0)}{2|\epsilon_F|+{\hbar\omega_{\bm \Gamma}^{static}(0)}}
\right|
\end{equation}
where $\alpha_{\bm \Gamma} = \frac{2A_0\langle g^2_{\bm \Gamma}\rangle_{\rm F}}{\pi
\hbar^2 v^2_F}$, $A_0 = 5.23$~\AA$^2$~ is the equilibrium unit cell area.
In this case, the energy shift diverges logarithmically when the magnitude
of the Fermi energy becomes the half of the phonon energy and 
increases in proportion to the Fermi energy for $|\epsilon_F| > \hbar\omega_0^{static}/2$
as long as the effect of the charge doping can be considered as a perturbation.
Beyond this,
one can easily notice a considerable change in $\alpha_{\bm \Gamma}$ and 
$\omega_{\bm \Gamma}^{static}$ with charge doping.

In order to calculate the dynamic frequency shift of Raman G band
 for graphene layers, we need    
$\alpha_{\bm \Gamma}$ for different layers. This is calculated using ab-initio
density functional theory and results are given in Table I. 
Further, we find that the static zero-doping frequencies 
$\omega_{\bf K}^{static} (0) = 1301.7$~cm$^{-1}$
and $\omega_{\bm \Gamma}^{static} (0) = 1553.3$~cm$^{-1}$ do not change with
the number of graphene layers, n.
We plot the dynamic frequency shift of
the Raman G band as a function of the charge doping 
for different graphene layers in Figure 3(a). One can readily notice (from
Figure 2) that for 
small doping (around
$0.4 \times
 10^{-13}$~ $cm^{-2}$ ) expected in zero-biased unintentionally doped graphene on a substrate, 
 the static frequency shift is very small
and can be treated as a correction ( $< 1$~ $cm^{-1}$ ) with 
respect to the dynamic one. For such a low doping 
concentration, the dynamic shift shows a linear dependence on 1/n (Figure 3(b)).
Interestingly, this is very similar to the 
the experimentally 
observed~\cite{r24} dependence of the 
the Raman G band position on the number of layers.
We, therefore, suggest that the frequency shift of the Raman
G mode on the number of layers is dominantly due to the unnitentional charge doping
which can be understood based on the dynamic effects. 
\begin{table} [h]
\begin{ruledtabular}
\begin{tabular}{ccccc}
\multicolumn{1}{c}{Number of}& \multicolumn{2}{c}{$\alpha$}& 
\multicolumn{2}{c}{$\gamma [T=0K]$ $(cm^{-1})$} \\

Layers, n& $\alpha_{\bm \Gamma}$&  $\alpha_{\bf K}$& $\gamma_{\bm \Gamma}$& 
$\gamma_{\bf K}$ \\
\hline
1 &   $4.441 \times 10^{-3}$ &   $13.33 \times 10^{-3}$ &    11.70 & 8.10  \\
2 &   $2.223 \times 10^{-3}$ &   $6.68 \times 10^{-3}$ &    11.56 & 8.04 \\
3 &   $1.457 \times 10^{-3}$ &                          &     11.43      \\
4 &   $1.085 \times 10^{-3}$ &                          &     11.32     \\
5 &   $0.889 \times 10^{-3}$ &                          &     11.23     \\
\end{tabular}
\end{ruledtabular}
\caption{Electron-phonon coupling strength, $\alpha$ and phonon-line widths, $\gamma$
$( cm^{-1} )$ at
${\bm \Gamma}$  and ${\bf K}$ for graphene layers. The values are computed
using 
density functional theory. $\alpha$ decreases very fast with increasing 
number of layers while the change in $\gamma$ is very small.
}
\label{tabella2}
\end{table}

Moreover, our DFT calculations (Table I) show that the phonon linewidth 
changes very little with the number of graphene layers. This result
is expected on physical grounds. In graphene layers,
the phonon linewidth due to the electron phonon coupling  
obtained from the
imaginary part of the phonon self-energy (or equivalently from the
Fermi Golden rule) is~\cite{r17,r7}
\begin{eqnarray}
&&\left(\gamma_{\bm \Gamma}\right)_n =
\frac{\pi \left(\alpha_{\bm \Gamma}\right)_n n
\hbar\omega^{static}_0}{2} \nonumber \\ 
&&\left[f\left( -{\hbar\omega^{static}_0}-2\epsilon_F\right)
-f\left( {\hbar\omega^{static}_0}-2\epsilon_F\right)\right] 
\label{eq7}
\end{eqnarray}
where n is the number of graphene layers. We find that 
$\left(\alpha_{\bm \Gamma}\right)_n n$
is almost constant with the number of layers, leading to 
a very small variation of phonon line width with number of layers.  
At zero temperature, we get
$\gamma($T=0K$)=\frac{\pi \left(\alpha_{\bm \Gamma}\right)_n n \hbar\omega_0^{static}}{2}~
 \Theta(\hbar\omega^{static}_0 - 2|\epsilon_F|)$.
It implies that the broadening is nonzero only for
$|\epsilon_F|<\hbar\omega^{static}_0/2$ when
the scattering process is allowed
by the Pauli exclusion principle.

\begin{figure} [h]
\centerline{\includegraphics[width=85mm]{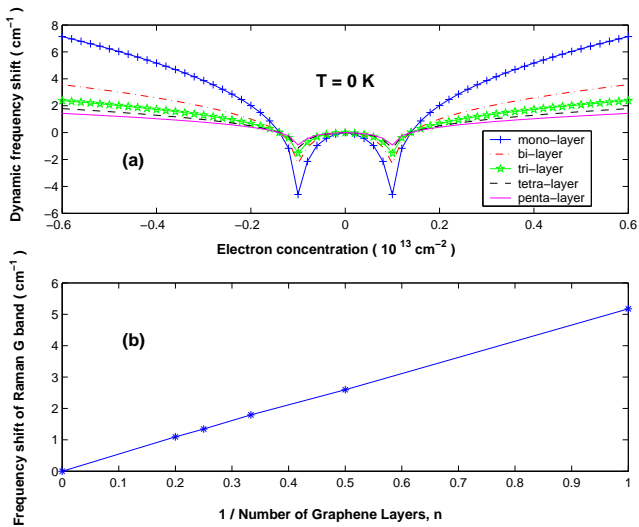}}
\caption{(Color online) (a) Dynamic frequency shift with charge doping in
graphene multi-layers and (b) Frequency shift of Raman G band
as a function of the number of layers.
}
\label{fig3}
\end{figure}

Finally, we discuss the effect of the charge doping on the 
${\bf K}$-point phonon. Recent experimental observation~\cite{r8} indicates
that  although the width and the peak position of the Raman 
G and $D^*$ bands ( second order of ${\bf {K}}$  point phonons ) 
exhibit similar doping dependence, the 
magnitudes of the changes are only around 10 percent of the
G band. This small dependence suggests that for the large wave vector 
phonon (${\bf K}$ point phonon) there is a cancellation
of the large inter-band and intra-band contribution~\cite{r25} in the
dynamic response and the static DFT frequency shift (see Figure 2)    
plays the major role.
We are also studying the layer dependence of the ${\bf K}$-point frequency 
shift in doped graphene multi-layers and that will be reported elsewhere
very soon.

In conclusion, we have computed, from first-principles, the frequency shift
of the A$'_1$-${\bf K}$ phonon (Raman D band) in monolayer graphene, as a
function of the charge doping. A detailed DFT study on the electron-phonon
coupling and the phonon line width of E$_{2g}$-${\bm \Gamma}$
and A$'_1$-${\bf K}$ phonons have also been performed for
graphene multi-layers. Furthermore, we explain the experimentally observed
'1/(Number of Layers)' behaviour of the
Raman G band position after including the dynamic response treated
within the time dependent perturbation theory.

S.K.S. gratefully acknowledges the financial support of Research
Fellowship from Council of Scientific and Industrial Research (India)
under Award Number 9/79(913)/2002-EMR-I. A.K.S. thanks Department of Science 
and Technology (India) for financial support.
We also thank Prof. Prabal K Maiti for providing access to his computer resources.


\begin{thebibliography}{99}

\bibitem{r1}
K.S. Novoselov {\it et al.} Science {\bf 306}, 666 (2004).

\bibitem{r2}
Y. Zhang et al, Appl. Phys. Lett. \textbf{86}, 073104 (2005).

\bibitem{r3}
K.S. Novoselov {\it et al.} Nature {\bf 438}, 197 (2005).

\bibitem{r4}
Y. Zhang, Y.W. Tan, H.L. Stormer, and P. Kim,  Nature 
{\bf 438}, 201 (2005).

\bibitem{r5}
K. S. Novoselov et al., Proc. Natl. Acad. Sci. USA
  \textbf{102}, 10451 (2005).

\bibitem{r6}
K. S. Novoselov et al., Nature Physics \textbf{2}, 177 (2006).

\bibitem{r7}
Simone Pisana {\it et al.} cond-mat/0611714 (2006).

\bibitem{r8}
J. Yan {\it et al.} cond-mat/0612634 (2006).


\bibitem{r9}
S. Piscanec, M. Lazzeri, F. Mauri, A.C. Ferrari, and J. Robertson,
Phys. Rev. Lett. {\bf 93}, 185503 (2004).

\bibitem{r10}
P. Hohenberg, W. Kohn Phys. Rev. {\bf 136}, B864 (1964);
W. Kohn and L.J. Sham Phys. Rev. {\bf 140}, A1133 (1965).

\bibitem{r11}
S. Baroni, S. de Gironcoli, A. Dal Corso, and P. Giannozzi,
http://www.pwscf.org.

\bibitem{r12}
J.P. Perdew, K. Burke, and M. Ernzerhof
Phys. Rev. Lett. {\bf 77}, 3865 (1996).


\bibitem{r13}
D. Vanderbilt, Phys. Rev. B {\bf 41}, 7892 (1990).

\bibitem{r14}
M. Methfessel and A.T. Paxton Phys. Rev. B {\bf 40}, 3616 (1989).

\bibitem{r15} 
http://www.library.cornell.edu/nr/bookcpdf/c10-7.pdf.

\bibitem{r16}
S. Baroni, S. de Gironcoli, A. Dal Corso, and P. Giannozzi,
Rev. Mod. Phys. {\bf 73}, 515 (2001).

\bibitem{r17}
M. Lazzeri and F. Mauri, Phys. Rev. Lett. {\bf 97}, 266407 (2006). 

\bibitem{r18}
T. Ohta {\it et al.} cond-mat/0612173 (2006). 

\bibitem{r19}
C. Thomsen and S. Reich, Phys. Rev. Lett. {\bf 85}, 5214 (2000).

\bibitem{r20}
A.C. Ferrari {\it et al.} Phys. Rev. Lett. {\bf 97}, 187401 (2006). 

\bibitem{r21}
D. Graf {\it et al.} Nano Lett., {\bf 7} (2), 238 -242, 2007.

\bibitem{r22}
For example, see, G. D. Mahan, Many Particle Physics, Third Edition, Page 140.

\bibitem{r23}
M. Lazzeri {\it et al.}
Phys. Rev. B {\bf 73}, 155426 (2006).

\bibitem{r24}
A. Gupta {\it et al.} cond-mat/0606593 (2006).

\bibitem{r25}
A. H. Castro Neto and Francisco Guinea, Phys. Rev. B {\bf 75}, 045404 (2007).






\end{thebibliography}
\end{document}